\documentclass{PoS}

\title{Physics with Nuclei at an Electron Ion Collider}

\ShortTitle{Nuclei at EIC}

\author{W. K. Brooks\\
	Institute for Advanced Studies in Physics and Engineering\\
	Valpara\'iso Center for Science and Technology\\
		Department of Physics, Universidad T\'ecnica Federico Santa Mar\'ia, Valpara\'iso, Chile\\
        E-mail: \email{william.brooks@usm.cl}}

\abstract{Lepto-nuclear colliders offer unique experimental opportunities to probe QCD in an extended medium. Of the many possibilities, three experiments are described here that are clearly feasible and of high scientific importance. First, a direct measurement of the saturation scale is possible using the broadening of the transverse momentum distribution of hadrons produced in semi-inclusive DIS. This connection to saturation physics will provide a quantitative measure of the degree to which this fundamental QCD phenomenon is occurring, and has important consequences in other high-energy scattering studies. Second, the measurement of quark energy loss is feasible, either as a direct measurement at a lower-energy EIC, or as an indirect measurement using hadron attenuation at higher energies. Such a measurement will offer the first experimental validation of the energy independence of this process, which is a consequence of the QCD analog of the LPM effect in QED. The third experiment is to determine the mechanisms of hadronization using the nucleus as a spatial analyzer, intercomparing hadron attenuation for nuclei of a range of sizes. The extended reach of the EIC in energy will allow probes of these mechanisms in the crucial high-z region for the heaviest nuclei, and will permit study of hadronization in heavy quark meson and baryon systems.
      }

\FullConference{XVIII International Workshop on Deep-Inelastic Scattering and Related Subjects, DIS 2010\\
		April 19-23, 2010\\
		Firenze, Italy}

\begin{document}
\section{Introduction}
The experimental exploration of QCD in DIS has historically focused on measurements with the proton, due to the relative simplicity of interpreting the results. The feasibility of semi-inclusive measurements, due in part to technical advances of the past two decades in accelerator and detector technology, has brought new theoretical ideas and opened up new subfields. However, some important aspects of QCD remain permanently inaccessible to measurements on proton targets. These important aspects are now being explored through the use of nuclear targets in semi-inclusive DIS, and motivate the focus on nuclei in future studies at an Electron-Ion Collider.

One of the principal advantages of nuclear targets is the higher resolving power made available by the multiple interactions of hadrons and partons passing through the spatially extended QCD system. Systematic studies with nuclei of varying size are revealing space-time properties of the strong interaction, such as the time scales and microscopic mechanisms of hadronization \cite{brooks09}. Experiments with the heaviest nuclei at higher energies will give access to the gluonic properties of matter, such as the gluon saturation that is a fundamental prediction of QCD \cite{levin1}. The power and precision of DIS on strongly interacting systems of extended size provides an important complement to the studies of these systems at higher densities and temperatures as probed in heavy-ion collisions \cite{majumder1}. To the extent that parton propagation through cold systems can be quantified and understood, it can be used as a tool to probe the hot and dense systems \cite{accardi1}. This provides a bridge that unites two historically disconnected subfields of  strong interaction studies. Further, it can be argued that the energy independence of the energy loss of quarks in cold nuclei is connected to QCD factorization \cite{brodsky1}, and also that the breakdown of factorization at large $x_F$ can be best examined in nuclei due to the higher resolving power available from multiple interactions in the nucleus \cite{kopeliovich1}. These topics are briefly discussed in the following sections, with emphasis on experimental possibilities at an Electron-Ion Collider.

\section{Determining the Saturation Scale from $p_T$ broadening}
Parton saturation is a topic of high current interest in those communities which study QCD at high energies \cite{rezaeian1}. The physical concept underlying this subject is that at high energies, i.e. at increasing $y = ln(1/x_B)$, parton densities grow rapidly, particularly for gluons. However, this growth, which is connected to the increase in total cross section with energy, cannot continue unbounded without eventually violating unitarity, and thus a saturation in the parton densities is anticipated. This expectation has a number of implications, such as the potential applicability of weak coupling techniques at sufficiently high $y$. For a review see \cite{cgc}. 

The saturation scale is essentially the characteristic scale of the distribution of the gluon transverse momentum (squared) at high energies. Within well-defined approximations, the saturation scale can be related to the gauge-invariant Transverse-Momentum-Dependent (TMD) gluon distribution in a nucleon embedded in a nucleus \cite{wang1}, connecting to the saturation scale via a dipole approximation. Within a dipole picture, the saturation momentum has been shown to be identically equal to the transverse momentum broadening from semi-inclusive DIS in a nucleus \cite {kopeliovich2}:

\begin{eqnarray}
Q_{sat}^2(b,E) = \Delta  p_T^2(b,E)
\end{eqnarray}

\noindent where $b$ is the impact parameter and $E$ is the energy of the parton propagating through the medium.

The physical origin of the broadening is the interaction of a propagating parton with the transverse gluonic field in the medium through gluon bremsstrahlung. The probability of gluon radiation is proportional to the gluonic parton density of the medium, and thus $p_T$ broadening is a direct measure of the saturation phenomenon.

The value of $\Delta p_T^2$ has been measured in a small number of experiments where the lab-frame parton energies range from 2 GeV to 270 GeV \cite{brooks09,kopeliovich2,hermes1}. Interpretation of the observed hadron broadening $\Delta p_T^2$ in terms of parton broadening $\Delta k_T^2$ requires taking into account the nature of the propagating entity (quark, gluon, photon, dilepton), the hadron species measured in the final state, and kinematical factors. The magnitude of hadron broadening in these experiments has ranged from $\Delta p_T^2 \approx 0.02 ~GeV^2$ for the lightest nuclei at the lowest energies to $\Delta p_T^2 \approx 0.5 ~GeV^2 $ for the heaviest nuclei at the highest energies. The small size of the effect and the rather slow growth with parton energy has implications for the experimental determination of broadening at an Electron Ion Collider using semi-inclusive DIS. On the one hand, the saturation phenomenon is clearly most relevant at highest energies, and the broadening signal is largest there. On the other hand, the experimental determination of this relatively small quantity places constraints on the angular resolution and momentum resolution of the measured hadron. Initial estimates indicate that adequate resolution is achievable given the EIC parameters currently under discussion for any near-term facility. Further, it is necessary to have some form of particle identification for the final state hadron, either with direct identification (e.g., a RICH detector) or through the analysis of invariant mass peaks, in which case the sensitivity to the broadening also depends on the characteristics of the combinatoric backgrounds. An additional analysis complication for $x_B<0.1$ is the dominance of dijet production. This process divides the available energy of the virtual photon $\nu$ between two quark jets and thus $z_h=E_{hadron}/\nu$ can no longer be uniquely interpreted as the fraction of the initial quark's energy carried by the final state hadron, as it can for $x>0.1$. However, the translation between the broadening observed in dijets and the underlying partonic broadening is still possible based on the known QED distribution amplitudes for a virtual photon to convert into a $q \bar{q}$ pair with fractional momenta $\alpha$ and $1-\alpha$ \cite{kopeliovich2}. Thus, with sufficient experimental resolution and particle identification, semi-inclusive DIS at an EIC can be used for precision determinations of the QCD saturation scale.

\section{Estimating Quark Energy Loss in Cold Nuclear Matter}

Quarks that propagate through strongly interacting systems lose energy through gluon radiation, as discussed above, and to a lesser extent through recoil losses in elastic scattering. The latter, which is nearly negligible for light quarks in cold matter, becomes more important for heavy quarks, and can also be more significant in hot dense matter. 
Theoretical work on quark energy loss in QCD has been ongoing for more than two decades \cite{accardi2,majumder2}, often with a focus on detecting and understanding the formation of hot dense matter in heavy ion collisions. The subject exhibits many complexities as well as parallels with analogous processes in QED.

Quark energy loss in arbitrarily thick strongly interacting media is characterized by the existence of a critical length $L_{crit} \propto \sqrt{E}$ that separates two regions; for parton energies of a few GeV, $L_{crit}$ is believed to be comparable to the dimensions of heavy nuclei. For pathlengths less than $L_{crit}$, the total energy loss $\Delta E$ is proportional to the square of the the pathlength and independent of energy, while for pathlengths greater than $L_{crit}$, $\Delta E$ is linear in pathlength and depends on the quark energy: 
\begin{eqnarray}
\Delta E \propto L^2, ~~~~~~~~L<L_{crit}; ~~~~~~~~~~\Delta E \propto L\sqrt{E}, ~~~~~~~~L>L_{crit}
\end{eqnarray}
With the span of energies available at a lower-energy EIC it should be possible to study the behavior of the energy loss near the critical length, and in particular to unambiguously determine its value for the interesting higher-energy region $L<<L_{crit}$, where the analog of the QED LPM effect is expected to occur. A second advantage of this region is that the production length \cite{kopeliovich3} can be significantly larger than nuclear dimensions, minimizing the effect of hadronization within the nuclear medium and removing ambiguity concerning the quark path length within the medium.

In cold nuclear matter and for nuclei of finite size, by all estimates the radiative energy loss of light quarks is at the level of a few hundred MeV per femptometer of path length. This constrains the possibilities for a direct measurement at high energies; for example, an energy loss of 1-2 GeV for the largest nucleus would imply that for a 100 GeV hadron the experimental sensitivity in terms of energy resolution should be at the 1-2\% level. Taken together with the requirement of unfolding the dijet distribution for $x_B<0.1$ as discussed in the previous section, it seems impractical to consider the direct method at high energies, although deconvolution of hadron attenuation will access this information. However, for lower hadron energies and $x_B>0.1$, a direct measurement is practical and would be of very high scientific interest since the model dependence would be minimal in that case.

\section{Hadronization Mechanisms from Hadron Attenuation}

Experimental studies over the past decade have made new inroads into understanding the mechanisms of hadronization using semi-inclusive DIS on nuclear targets. First HERMES at DESY, and now CLAS at Jefferson Lab, are providing new data and new insights on how hadrons form starting from propagating quarks, using the nuclear systems as spatial analyzers with known characteristics. These experiments have spurred much theoretical activity and progress. Another wave of new data will arrive following the startup of the 12 GeV upgrade at Jefferson Lab from experiment E12-06-117, "Quark Propagation and Hadron Formation," which will probe nuclear targets with three orders of magnitude more integrated luminosity than the HERMES program, using 11 GeV electrons. This program will thoroughly explore the low-energy phenomena associated with meson and baryon formation, including rare processes accessible through the high luminosity, such as $\phi$ meson formation. However, the behavior for $\nu>9$ GeV will not be accessible; this region is important for understanding the energy dependence of hadron formation, particularly for heavy nuclei. Further, access to baryon formation studies will be limited to light quark baryons, while much interesting physics could be learned from heavy quarks. An EIC at lower energies and  highest luminosity could address the full energy dependence of hadron formation and the formation patterns of heavy quark systems. At higher energies, studies of this type could test the prediction of the breakdown of factorization at high $z$ (large $x_F$, \cite{kopeliovich1}) and study its systematic behavior. A detailed understanding of this phenomenon could be important for interpretation of high-energy interactions in many different contexts.

\section{Conclusion}
Three experimental opportunities have been described. These take advantage of the unique energy range and good luminosity of the EIC. The three lines of investigation are: direct measurement of the saturation scale, determination of quark energy loss, and investigation of the mechanisms of hadronization. These exciting topics are important both for their fundamental impact on understanding QCD as well as for their relevance to investigations at high energies worldwide.

\end{document}